\documentclass[conference]{IEEEtran}
\IEEEoverridecommandlockouts
\usepackage{cite}
\usepackage{amsmath,amssymb,amsfonts}
\usepackage{algorithmic}
\usepackage{graphicx}
\usepackage{textcomp}
\usepackage{xcolor}
\usepackage{amssymb}
\usepackage[colorlinks=true, citecolor=black, urlcolor=black, linkcolor=black]{hyperref}

\def\BibTeX{{\rm B\kern-.05em{\sc i\kern-.025em b}\kern-.08em
    T\kern-.1667em\lower.7ex\hbox{E}\kern-.125emX}}

\usepackage{mathtools}

\begin{document}

\title{\textit{Forte}: An Interactive Visual Analytic Tool for Trust-Augmented Net Load Forecasting\\
\thanks{Parts of this work were supported by the U.S. Department of Energy Solar Energy Technologies Office and the Office of Electricity Sensors Program, and performed jointly at the Pacific Northwest National Laboratory under Contract DE-AC05-76RL01830 and at the Lawrence Livermore National Laboratory under Contract DE-AC52-07NA27344).}
}


\author{\IEEEauthorblockN{Kaustav Bhattacharjee\IEEEauthorrefmark{1}, Soumya Kundu\IEEEauthorrefmark{2}, Indrasis Chakraborty\IEEEauthorrefmark{3} and Aritra Dasgupta\IEEEauthorrefmark{1} }

\IEEEauthorblockA{\IEEEauthorrefmark{1}\textit{Department of Data Science}, New Jersey Institute of Technology, USA. Email: \{kb526,\,aritra.dasgupta\}@njit.edu} 
\IEEEauthorblockA{\IEEEauthorrefmark{2}\textit{Optimization and Control Group}, Pacific Northwest National Laboratory, USA. Email: soumya.kundu@pnnl.gov}
\IEEEauthorblockA{\IEEEauthorrefmark{3}\textit{Center for Applied Scientific Computing}, Lawrence Livermore National Laboratory, USA. Email: chakraborty3@llnl.gov} 
}

\maketitle

\begin{abstract}
Accurate net load forecasting is vital for energy planning, aiding decisions on trade and load distribution. However, assessing the performance of forecasting models across diverse input variables, like temperature and humidity, remains challenging, particularly for eliciting a high degree of trust in the model outcomes. In this context, there is a growing need for data-driven technological interventions to aid scientists in comprehending how models react to both noisy and clean input variables, thus shedding light on complex behaviors and fostering confidence in the outcomes. In this paper, we present \texttt{Forte}, a visual analytics-based application to explore deep probabilistic net load forecasting models across various input variables and understand the error rates for different scenarios. With carefully designed visual interventions, this web-based interface empowers scientists to derive insights about model performance by simulating diverse scenarios, facilitating an informed decision-making process. We discuss observations made using \texttt{Forte} and demonstrate the effectiveness of visualization techniques to provide valuable insights into the correlation between weather inputs and net load forecasts, ultimately advancing grid capabilities by improving trust in forecasting models.
\end{abstract}

\begin{IEEEkeywords}
visual analytics, AI/ML, net load forecasting
\end{IEEEkeywords}

\section{Introduction}
The net load of an electric grid can be defined as the difference between the total electricity demand and the electricity generation from behind-the-meter resources such as solar and other distributed generators~\cite{NetLoadDefinition}. It can vary based on various factors, including weather conditions and the time of the day. Accurate net load forecasting enables grid operators, policymakers, and energy providers to make informed decisions regarding energy trade, load distribution, and resource allocation. However, the proliferation of solar energy generation sources in residential settings has significantly impacted the performance of traditional net load forecasting models~\cite{SolarEnergyPenetration}.
%
We have collaborated with scientists who have developed a deep-learning model that produces probabilistic net load forecasts incorporating variables such as temperature, humidity, apparent power, and solar irradiance, achieving strong predictive performance and resilience in the face of missing data~\cite{sen2022kpf}. However, in order to improve trust in the model, these outputs need to be explored by domain experts, including scientists and grid operators. The model's performance may fluctuate due to seasonal variations in the input variables, and stakeholders must also assess its reliability in the face of noisy inputs mirroring real-life scenarios. Hence, the process is complex and time-consuming, prompting the need for an approach capable of performing these tasks and enhancing trust in the model's performance. Visual analytics can be instrumental here since prior research has shown that it can significantly enhance trust in model outputs during complex sense-making tasks~\cite{dasgupta2016familiarity}. In \cite{kandakatla2020towards}, it was argued that visual analytics would play a critical role in enabling trust-augmented artificial intelligence and machine learning (AI/ML) applications in energy sector.

In light of this, we collaborated with energy scientists to thoroughly investigate the model's performance across diverse time periods and input scenarios, gaining valuable insights into the evaluation tasks needed to comprehend the model's effectiveness. Building upon this experience, we performed a design study to develop a system aimed at facilitating stakeholders in efficiently performing these evaluation tasks. As a result, we developed a visual analytics-based application \texttt{Forte} that empowers users to gain an in-depth understanding of the model's performance, effectively leveraging data visualization techniques to aid informed decision-making in the realm of energy planning and grid operations.

Our application aims to provide a broad understanding of various aspects related to net load forecasting. First, it enables researchers and scientists in the energy domain to assess net load variability concerning input variables by comparing model forecasts with actual net load values across different time periods and seasons. They can gain insights into their impact on model performance by analyzing the effects of variables like temperature, humidity, and apparent power on net load forecasts. Second, \texttt{Forte} helps evaluate forecast errors with noisy inputs at different noise levels, thus providing information for improving the model's reliability and robustness in real-world scenarios. This visual analytics-based approach can empower scientists and grid operators to make data-driven decisions, enhancing trust and confidence in the net load forecasting model.

While prior research has primarily concentrated on developing interfaces during the model development process, tailored to aid model developers, our focus lies in the post-hoc evaluation of the model's performance, catering specifically to the needs of energy scientists and grid operators~\cite{lucas2021lumen, klaus2023visual}. Other works explore the performance of probabilistic net load forecasting models through different visualization charts but do not offer an integrated interactive interface~\cite{wang2017data, henriksen2022electrical}. Our visual analytic tool, \texttt{Forte} presents an integrated workflow that empowers users to explore net load variability and forecast error analysis concerning various input variables and scenarios, which makes it a novel approach in this domain.

In this paper, we first introduce our visual analytics-based application, \texttt{Forte}, developed in collaboration with energy scientists~(Section \ref{section:design}). Emphasizing the analytical goals and tasks, we also provide insights into the underlying design rationale. Subsequently, we present observations gleaned through our application that can potentially drive advancements in grid operations~(Section \ref{section:results}). Finally, we conclude with the lessons learned from this design study and how we incorporate them into our application~(Section \ref{section:conclusion}). Additionally, a short demonstration video is available at \href{https://bit.ly/forte_demo}{this link}. 

\section{Visual analytics-based design}\label{section:design}
\begin{figure*}
\begin{center}
\includegraphics[width=\textwidth]{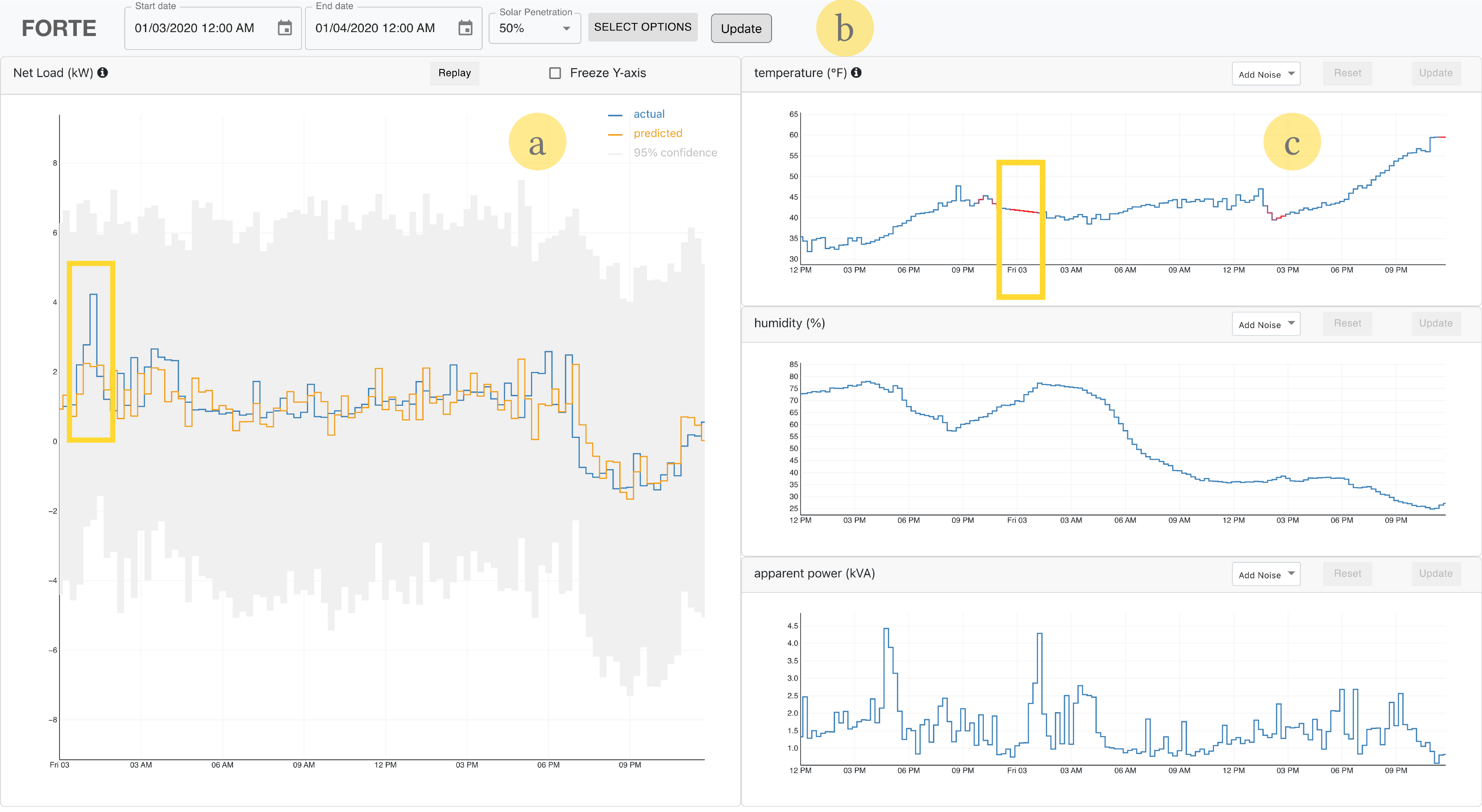}
 \caption{\label{figs:interface}\textbf{The interface for our net load forecasting visual analytic tool~(\texttt{Forte}):} (a) Our application facilitates the comparison of actual and predicted net load within the selected time frame and solar penetration levels as defined (b) through the Options Selection Area. Further, (c) the influence of various weather conditions on predictions can be explored via the Inputs View Area. The highlighted region shows instances of missing temperature data and resultant disagreement between predicted and actual net load within the same time period. These insights are valuable to grid operators as it allows them to review the data quality, evaluate its impact on model performance, and make recommendations for sensor/metering upgrade.}
\end{center}
\vspace{-6mm}
\end{figure*}
Our application \texttt{Forte} adopts a visual analytics-based design that integrates coordinated views aided with visual cues to show the various aspects of net load forecasting outcomes. It combines interactive visualization with performance metrics to instill greater trust in model outcomes, providing users with the flexibility to probe net load predictions as a function of input variables like temperature and humidity. In this section, we outline our \texttt{Forte}'s goals and tasks, followed by an explanation of our design rationale.
We aim to achieve the following visual analytic tasks via \texttt{Forte}:
\par \noindent \textit{\textbf{T1}: Understand actual net load and predictions across time periods and solar penetration levels:} The net load forecasting model is trained with data from varying solar penetration levels, while energy consumption fluctuates throughout the day and across seasons. Thus, this task involves comprehending the model's performance across diverse time spans and solar penetration levels.
\par \noindent \textit{\textbf{T2}: Explore the impact of input variables on net load prediction:} Different inputs, such as weather conditions, influence the forecasting model differently. Therefore, it becomes essential to grasp the effects of these input variables on the model across diverse time spans, which motivates this task.
\par \noindent \textit{\textbf{T3}: Augment missing data with background knowledge:} Real-world weather data frequently includes gaps due to various factors. However, a domain expert may possess insights into expected temperature or humidity for specific timeframes. This task involves empowering users of the application to adjust inputs and observe how these modifications influence the model's performance.
\par \noindent \textit{\textbf{T4}: Design experiments simulating different noisy input scenarios:} Noisy inputs can vary due to factors like noise levels, direction (bidirectional/uni-directional), and number of observations. This task involves crafting scenarios using these factors to simulate and explore noise effects.
\par \noindent \textit{\textbf{T5}: Assess model efficacy across different months and varying levels of noise:} The model's sensitivity to noisy inputs can differ among months and noise levels. This task involves studying how varying noise levels affect model performance across different months.

Next, we explain our application's design by outlining its various high-level goals and demonstrating how it accomplishes these tasks.

\subsection{Goal: Understand Net Load Forecasts w.r.t Input Variables}
Understanding the interplay between net load forecasts and input variables is essential for making informed decisions in energy planning and ensuring efficient grid operations. Towards this end, \texttt{Forte} integrates three essential components:

\textbf{\textit{Options Selection Area:}} As mentioned earlier, net load forecasts can fluctuate based on time periods and solar penetration levels. Accordingly, \texttt{Forte} offers these selections prominently at the top, within the Options Selection Area~(Figure~\ref{figs:interface}b). This area contains two date and time pickers, enabling users to specify their preferred observation timeframe. Currently, users can opt for any date within the year 2020, with the potential for expansion as further data is available. Additionally, users can choose solar penetration levels from 0\%, 20\%, 30\%, and 50\%. This area provides options for choosing different prediction horizons (15 minutes or 24 hours ahead) and input variables (temperature, humidity, apparent power, etc.) tailored to user preferences. Initially, a limited set of input variables is loaded to reduce visual clutter, allowing users to add more based on their choices.

\textbf{\textit{Net Load View Area:}} This component facilitates a direct comparison between the actual net load and the predicted net load for the chosen time period and solar penetration level, as selected within the Options Selection Area~\textbf{(T1)}. This visual representation employs a blue line to depict the actual net load and an orange line to depict the predicted net load~(Figure~\ref{figs:interface}a). 
The extent of proximity/overlap between these lines indicates the level of agreement between actual and predicted net load, reflecting a superior model performance. However, the degree of agreement can also be quantified by metrics like Mean Absolute Error~(MAE)~\cite{willmott2005advantages} and Mean Absolute Percentage Error~(MAPE)~\cite{tayman1999validity}, which are revealed by hovering on the icon button atop this area.

Since our net load forecasting model produces a probabilistic forecast, we additionally present the 95\% confidence interval for this forecast, indicated by a subtle, shaded grey area. This design choice was made to streamline the view by avoiding the introduction of two additional lines, effectively minimizing visual clutter within this area. When users modify the options, we noted that the Y-axis within this area might shift due to value changes, impeding the observation of variations across distinct time periods or solar penetration levels. This problem can be alleviated using the ``Freeze Y-axis" option, which, as the name suggests, freezes the Y-axis at the current values and plots the new values based on the frozen axis. Additionally, changes can be tracked using the Replay button, which showcases net load changes through a slower animation~($\thickapprox10$s).

\textbf{\textit{Inputs View Area:}} Located on the right-hand side of the application, the Inputs View Area displays the selected inputs (as selected from the Options Selection Area) and their respective values during the chosen time period~(Figure~\ref{figs:interface}c). It also shows some of the historical data used while generating the forecast for this period. This visualization aids in establishing correlations between weather data and the agreement/disagreement observed between the actual and predicted net load, thereby impacting model performance~\textbf{(T2)}.

Nonetheless, weather data might feature gaps for specific time spans, which are addressed through linear interpolation connecting the nearest available data points. These interpolated points are indicated in red, and users have the flexibility to drag and adjust them based on their expertise~\textbf{(T3)}. The data quality, denoting the percentage of missing data, can be accessed by hovering over the icon button atop each input variable. However, if the users do not trust the quality of the available weather data, they can apply a uniform noise of 5\% or 10\% via the ``Add Noise" option for each input variable. All these changes are reflected on the Net Load View Area once users hit the ``Update" button. Thus, \texttt{Forte} begins with simple visualization and default settings, easing the learning curve as users delve into advanced features.
\subsection{Goal: Compare model performance w.r.t Noisy Inputs}
During our investigation into the influence of input variables on net load prediction, we noticed that the model's responses varied with different noise levels. Hence, we developed a separate linked page with the following components.

\begin{figure*}
\begin{center}
\includegraphics[width=\textwidth]{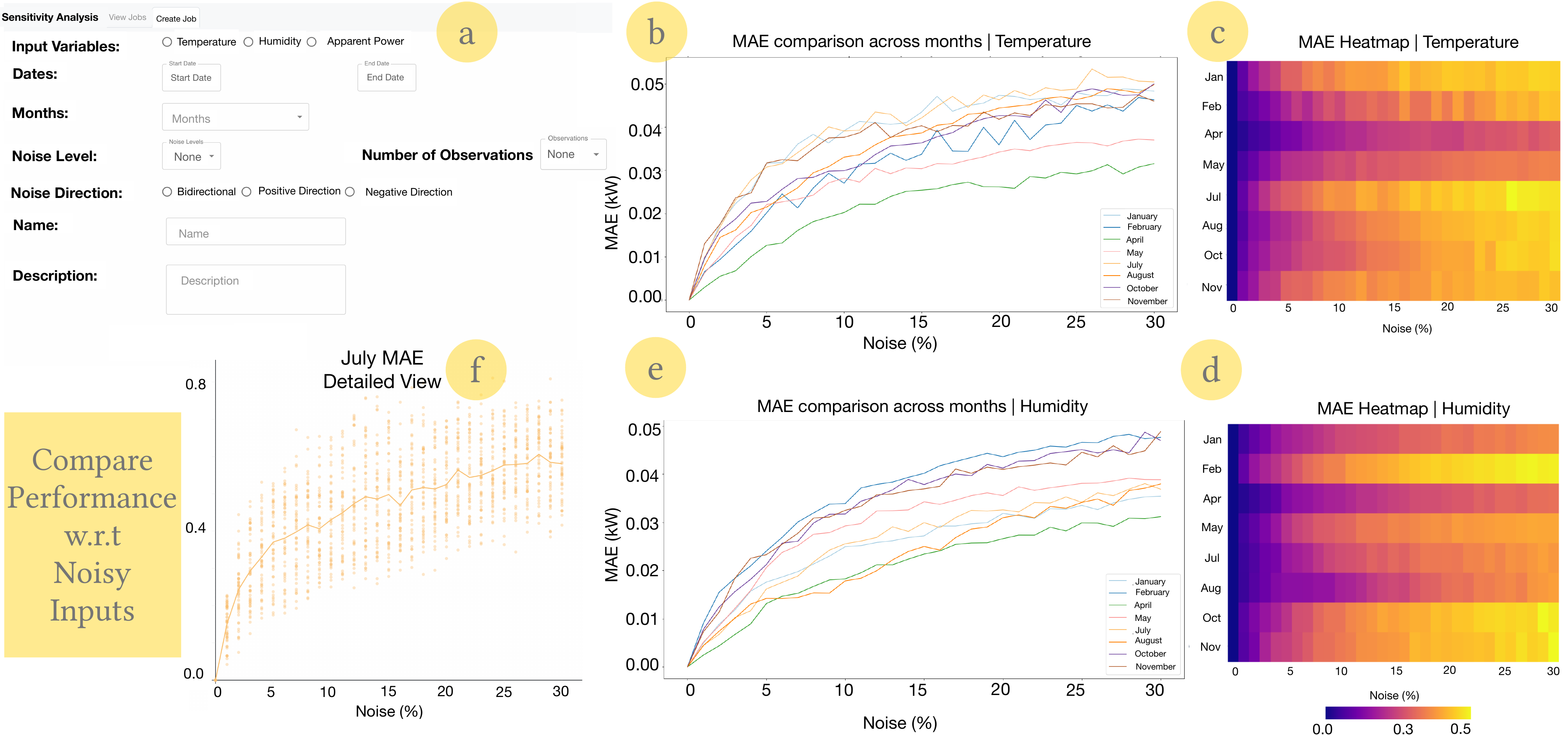}
 \caption{\label{figs:noisy_inputs}\textbf{Experimental Results:} (a) Our application \texttt{Forte} enables the design of experiments through the creation of noisy inputs using various factors; and the results (error rates) can be cross-compared across various months for both the input variables of (b, c) temperature and (d, e) humidity; (f) with the option to view detailed observations for each month. These insights generated through \texttt{Forte} are valuable to the user (a grid operator) to not only reveal the underlying dependence of the model outcome (net load prediction) on different input weather conditions but also better prepare ahead of any impending weather events (e.g., heat/cold wave).}
\end{center}
\vspace{-6mm}
\end{figure*}

\textbf{\textit{Experiment Design Area:}} Users can generate simulated noisy inputs for varying dates spanning multiple months and a specific input variable. This area empowers users to select their preferred input variable (temperature, humidity, apparent power), set start and end dates, and designate desired months for introducing noise~\textbf{(T4)}. The area also offers the flexibility to add or subtract a uniform noise (ranging from 1\% to 30\%)  from the original inputs or use a combination thereof~(Figure~\ref{figs:noisy_inputs}a). As the noise is uniformly distributed, the Experiment Design Area accommodates multiple observations/forecasts with identical inputs. Finally, users can add a name and short description for future reference, and our application will show an estimated time for completing this experiment. This experiment-based architecture enables \texttt{Forte} to manage computational overload efficiently.

\textbf{\textit{Experiments View Area:}} Once the designed experiments are complete, those are available for the user's perusal. Users can select any of the completed experiments/jobs from the left-hand side navigation bar. Each experiment initially displays two line charts depicting the error metrics MAE and MAPE~(Figure~\ref{figs:noisy_inputs}b,~\ref{figs:noisy_inputs}e). These charts comprise lines corresponding to the months chosen during the experiment design. Each line illustrates the deviation in error metrics from their baseline values (established at 0\% noise or no noise)~\textbf{(T5)}. We offer an alternative visualization in the form of a heatmap, illustrating the deviation in MAE from the baseline values for each month. Based on initial feedback, users found this heatmap particularly useful for comparing the model's sensitivity across different months~(Figure~\ref{figs:noisy_inputs}c,~\ref{figs:noisy_inputs}d).
In addition to this, users may want to explore the error rates for each month. Hence, we also include two scatterplots for each month (for each of the error metrics), which show the error rates for each of the observations~(Figure~\ref{figs:noisy_inputs}f). This scatterplot is then augmented with a line showing the average error rate for that month across different noise levels, mimicking the corresponding line in the first line chart. This view area helps to understand the model's performance when faced with noisy inputs and, in the process, improves trust in the model. \texttt{Forte} is primarily developed using React.js
and D3.js for the frontend, and Flask framework in Python for the backend. 

\section{Experimental Results}\label{section:results}

In this section, we showcase some outcomes from our application, emphasizing their possibility to enhance model training and potentially streamline grid efficiency. We illustrate this through a practical scenario involving a research scientist named Amy. Amy seeks to comprehend the influence of noisy inputs on the model and enhance trust in net load forecasts, consequently aiding effective grid operations planning.

Amy scrutinized net load predictions for January 3rd to 4th, 2020, at a 50\% solar penetration level through our application~\textbf{(T1)}. She noted a general alignment between predictions and the actual net load values, barring a deviation around 1 a.m. on Friday, January 3rd~(Figure~\ref{figs:interface}). Intrigued by this discrepancy, she used our application to delve into the temperature data for that period~\textbf{(T2)}. Here, Amy observed some missing data around the same time. The data around this timeframe underwent linear interpolation using the nearest available data, which Amy speculated might contribute to the discrepancy. To investigate, she interactively adjusted the line chart at this point, thereby updating the temperature values for that period~\textbf{(T3)}. This led to slight prediction variations, suggesting temperature's significance as an input variable to the model. As a further step, she introduced a uniform 5\% noise to all temperature values within the selected time period. Interestingly, this led to several changes in the predictions.

Now, Amy aimed to systematically comprehend the influence of noisy temperature values on the net load forecasting model. She devised an experiment by (arbitrarily) selecting the 3rd and the 4th days of various months in 2020 and introducing consistent bias/noise, ranging from 1\% to 30\%, to the recorded temperature values \textbf{(T4)}~(Figure~\ref{figs:noisy_inputs}a). Surprisingly, she observed no change in the error metrics. Her inference was that the model normalizes inputs before generating the predictions, thus explaining the similar outputs despite varying noisy inputs.
Subsequently, Amy replicated the experiment, introducing uniform noise to the temperature values. As an example, for a temperature of $60^{\circ}$F with 10\% added noise, the range of noisy input could span from $60^{\circ}$F to $66^{\circ}$F. Given the randomized nature of this experiment, she chose to replicate it across 50 observations/iterations, akin to repeated measures design~\cite{ellis1999repeated, sullivan2008repeated}. Having reviewed the findings of this experiment, she proceeded to repeat it over eight months (January, February, April, May, July, August, October, and November). Her observations unveiled that, although error rates were minimal, the model displayed heightened sensitivity to noisy data during January and July, across numerous noise levels—albeit with exceptions. In contrast, the model exhibited the least sensitivity during April and May~(Figure~\ref{figs:noisy_inputs}b and~\ref{figs:noisy_inputs}c). 
The observed variations in the model's sensitivity to noisy perturbations in temperature data across different months, can be attributed to the influence of seasonal weather variations on usage of electricity~\textbf{(T5)}. For example, typically the heating and cooling load -- which drives the residential energy demand -- typically peaks during the coldest (e.g., January) and the hottest (e.g., July) months, thereby ensuring heightened sensitivity of net load to temperature variations. 
In contrast, sensitivity of residential energy usage to temperature perturbations remain low in shoulder months (e.g., April and May) with typically milder weather. Given the potential impact of climate change on these results over time, it is imperative to use \texttt{Forte} to conduct further such experiments regularly.
Insights revealed from this experiment helped Amy gain confidence in the model. This also underscores the embedded learning process, as these insights serve as valuable resources for retraining the model to handle noisy scenarios better.

Subsequently, Amy sought to determine if humidity yielded similar effects on the model's performance. She initiated a parallel experiment focusing on humidity~(Figure~\ref{figs:noisy_inputs}d and ~\ref{figs:noisy_inputs}e).
Notably, her observations indicated heightened model sensitivity during February and October, with reduced sensitivity aligning with April—mirroring the earlier temperature findings.
Furthermore, she delved deeper into the predictions for each month, aiming to grasp the distribution of error metrics across noise levels within 50 observations~(Figure~\ref{figs:noisy_inputs}f). While noting the presence of outliers in these error metrics, Amy observed that the mean line of these observations effectively captured the trend across most months. On an overall assessment, Amy discerned that while error rates varied across different months, the model consistently demonstrated commendable performance, with notably low error rates difference from the baseline (\,$\approx$\,0.05\,kW MAE). We can thus conclude that our visual analytics tool \texttt{Forte} effectively enabled her to grasp the model's performance concerning diverse weather data and their noisy variants, thus improving her trust in this model.
\section{Conclusion}\label{section:conclusion}
The significance of accurate net load forecasting in energy planning and grid operations cannot be overstated. Hence, in this study, we explored net load forecasting, leveraging a collaborative approach with domain experts to develop a visual analytics-based application. 
By partnering with energy scientists, we not only identified critical evaluation tasks but also translated them into an intuitive interface that empowers users to make sense of complex model behaviors.

Throughout this endeavor, we gained valuable insights into the challenges posed by noisy inputs, seasonal variations, and the need to instill trust in forecasting models. The collaborative process exposed the complexities of real-world data analysis and emphasized the necessity for efficient, user-friendly tools that bridge the gap between model insights and actionable decisions. We can argue that our application is a first step towards this direction. As a next step, we plan to elucidate the input normalization process, add other error metrics, and incorporate economic planning and analysis to enable stakeholders to gauge the cost-benefit ratio and enhance trust in the net load forecasting model~\cite{sassone1978cost, alaqeel2019comprehensive}.

Looking ahead, there are multiple areas where we would like to enhance \texttt{Forte}. Incorporating more datasets spanning multiple years, expanding to additional weather conditions, automating aspects of the experiment design area, and even using transfer learning techniques
are all promising avenues for future exploration. As the energy landscape evolves, our application's adaptability will play a vital role in helping energy planners, grid operators, and policymakers navigate the complexities of net load forecasting. In conclusion, our collaborative effort has yielded a powerful tool that not only deepens our understanding of net load predictions but also lays the foundation for more informed and efficient energy planning decisions in the years to come.


\bibliographystyle{abbrv-doi}
\bibliography{bib}

\end{document}